\input harvmac.tex
\def\b{\beta^2}
\def\l{\lambda}

\lref\leggett{Caldeira, A.O.
and Leggett, A.J.: Influence
of dissipation on quantum tunneling in
macroscopic systems. Phys. Rev. Lett. 
{\bf 46}, 211-214 (1981)\semi
Caldeira, A.O.
and Leggett, A.J.:
Path integral approach to quantum
Brownian motion.
Physica  {\bf A121}, 587-616 (1983)}

\lref\Callan{Callan, C.G. and Thorlacius, L.:
Open string theory as dissipative quantum 
mechanics. Nucl. Phys. {\bf B329}, 117-138 (1990)}

\lref\Fisher{Fisher, M.P.A. and Zwerger, W.:
Quantum Brownian motion in a
periodic potential.  Phys. Rev.
{\bf B32}, 6190-6206 (1985)}

\lref\Va{Voros, A.: Spectral Zeta functions. 
Adv. Stud. Pure Math. {\bf 21}, 
327-358, (1992)} 

\lref\Vb{Voros, A.: Exact  quantization
condition for anharmonic oscillators 
(in one dimension). J. Phys. {\bf A27}, 4653-4661 (1994)} 

\lref\Vc{Voros, A.: Airy function (exact WKB results for potentials 
of odd degree). Saclay preprint T98/101 (1998), {\tt math-ph/9811001}}

\lref\Fen{Fendley, P.: Unpublished}

\lref\DT{Dorey, P. and Tateo, R.: Anharmonic oscillators, 
the thermodynamic Bethe ansatz, and nonlinear integral equations. 
Preprint DTP-98-81, ITFA 98-41 (1998), {\tt hep-th/9812211}}

\lref\BLZnon{Bazhanov, V.V., Lukyanov, S.L. and Zamolodchikov, A.B.:
On nonequilibrium states in  QFT model with boundary interaction.
Preprint RU-98-51 (1998), {\tt hep-th/9812091}
}

\lref\Baxn{Baxter, R.J.: Partition function of the eight-vertex 
lattice model, Ann. Physics {\bf 70}, 193--228 (1972)}

\lref\BLZZZZ{Bazhanov, V.V., Lukyanov, S.L. and Zamolodchikov, A.B.:
Integrable quantum field theories in finite volume: Excited state 
energies. 
Nucl. Phys. {\bf B489}, 487-531 (1997), {\tt hep-th/9607099}} 

\lref\DDV{Destri., C. and de Vega, H.J.:
Unified approach to thermodynamic Bethe Ansatz 
and finite size corrections
for lattice models and field theories. 
Nucl. Phys. {\bf B438}, 413-454 (1995)}

\lref\BLZ{Bazhanov, V.V., Lukyanov, S.L. and Zamolodchikov, A.B.:
Integrable structure of conformal field theory,
quantum KdV theory and
thermodynamic Bethe Ansatz.
Commun. Math. Phys. {\bf 177}, 381-398 (1996) \semi
Bazhanov, V.V., Lukyanov, S.L. and Zamolodchikov, A.B.:
Integrable structure of conformal field theory III. The Yang-Baxter
relation. Preprint RU-98-14 (1998), {\tt hep-th/9805008}}

\lref\BLZZ{Bazhanov, V.V., Lukyanov, S.L. and Zamolodchikov, A.B.:
Integrable structure of conformal field theory 
II. Q-operator and DDV equation. Commun. Math. Phys. {\bf190}, 247-278
(1997), {\tt   hep-th/9604044}\semi
Bazhanov, V.V., Lukyanov, S.L. and Zamolodchikov, A.B.:
Integrable structure of conformal field theory III. The Yang-Baxter
relation. Preprint RU-98-14 (1998), {\tt hep-th/9805008}}

\Title{\vbox{\baselineskip12pt\hbox{RU-99-01}
\hbox{hep-th/9812247}}}
{\vbox{\centerline{Spectral determinants for Schr\"odinger equation}
{\centerline{}} 
{\centerline{and Q-operators of Conformal Field Theory}}
\vskip3pt\centerline{  }}}
\centerline{Vladimir V. Bazhanov$^1$,
Sergei L. Lukyanov$^{2}$}
\centerline{ 
and Alexander B. Zamolodchikov$^{2}$ }
\centerline{ }
\centerline{$^1$Department of 
Theoretical Physics and Center of Mathematics}
\centerline{and its Applications, IAS, Australian National University, }
\centerline{Canberra, ACT 0200, Australia}
\centerline{ }
\centerline{$^2$Department of Physics and Astronomy,}
\centerline{Rutgers University, Piscataway, NJ 08855-049, USA}
\centerline{and}
\centerline{L.D. Landau Institute for Theoretical Physics,}
\centerline{Chernogolovka, 142432, Russia}
 
\centerline{}
\centerline{}

\centerline{{\bf Abstract}}
\centerline{}
Relation between the vacuum eigenvalues of CFT ${\bf Q}$-operators and
spectral determinants of one-dimensional Schr\"odinger 
operator with
homogeneous potential, recently conjectured by Dorey and Tateo for
special value of Virasoro 
vacuum parameter $p$, is proven to hold, with suitable
modification of the Schr\"odinger operator, 
for all values of $p$.

\centerline{}
 
\Date{December, 98}
\vfil
\eject
\vglue 1cm

In recent remarkable paper\ \DT\ a novel relation was observed
between the vacuum eigenvalues of so-called ${\bf Q}$-operators
introduced in\ \BLZZ\ and the spectral characteristics of the 
Schr\"odinger
equation
\eqn\anharm{
\Big\{-\partial_x^2+|x|^{2\alpha}\, \Big\}\, \Psi(x)=E\, \Psi(x)\ .}
Namely, for special value of the vacuum 
parameter $p$ (see below) the vacuum
eigenvalues of the operators ${\bf Q}_{+}$ and ${\bf Q}_{-}$ essentially
coincide with the spectral determinants associated with the odd and even
sectors of \anharm, respectively. In this note we show that  a similar
relation, with an appropriately generalized spectral problem \anharm,
holds for all values of the vacuum parameter $p$.

The ${\bf Q}$ operators were constructed in\ \BLZZ\ in our attempt to
understand $c<1$ CFT as completely integrable theory; they appear to be
the CFT versions of Baxter's $Q$-matrix which plays most important role 
in his famous solution of the eight-vertex
model\ \Baxn. The ${\bf Q}$-operators of
\BLZZ\ are actually the operator functions ${\bf Q}_{\pm}(\lambda)$,
where $\lambda$ is a complex parameter. These operators act
in Virasoro module with the highest weight $\Delta$, this weight and the 
Virasoro central charge $c$ being conveniently parameterized as 
$\Delta=({p/\beta})^2+{(c-1)/24}$ and $c=1-6(\beta-\beta^{-1})^2$, with
$0 < \beta^2 \leq 1$. The highest weight state $\mid p\, \rangle$ (the
Virasoro vacuum) is an eigenstate of ${\bf Q}_{\pm}(\lambda)$ and we use
the notation 
\eqn\zkasiu{\lambda^{\pm 2\pi i p /\beta^2}\,A_{\pm}(\lambda, p) =
\langle\, p \mid {\bf Q}_{\pm}(\lambda) \mid p\, \rangle}
for the
corresponding eigenvalues. These eigenvalues deserve most detailed
study, for various reasons (see \refs{\BLZZ, \BLZnon}). 
Let us mention here
some of their properties relevant to the present discussion (details and
derivations can be found in\ \refs{\BLZZ,\BLZnon}). 

(i) ${A}_{\pm}(\lambda, p)$ 
are entire functions of the variable $\lambda^2$ with known asymptotic 
behavior  
\eqn\asympt{
\log { A}_{\pm}(\lambda, p)\simeq M (-\lambda^2)^{1\over2-2\b},\qquad
|\lambda^2|\to\infty,\qquad \arg(-\lambda^2)<\pi\ ,}
where
$$M={\Gamma\big({\beta^2\over 2-2\beta^2}\big)\,
\Gamma\big({1-2\beta^2\over 2-2\beta^2}\big)\over \sqrt{\pi}}\ 
\big(\Gamma(1-\beta^2)\big)^{{1\over 1-\beta^2}}\ .$$

(ii) $A_{+}(\lambda,p)$ is meromorphic function of $p$, analytic in the
half-plane $\Re e(2 p) > -\b$, and $A_{-}(\lambda, p) = A_{+}(\lambda,
-p)$.
The coefficients $a_n(p)$ of the power series expansion
\eqn\logA{\log A_{+}(\lambda, p)=
-\sum_{n=1}^{\infty}\, \ a_n(p)\ \lambda^{2 n}  }
exhibit the following asymptotic behavior
\eqn\osoeuiu{a_n(p)\sim p^{1-2 n+2 n \beta^2}\ \ \
\ \ \  {\rm as}\  \ p\to\infty }
in the half-plane $\Re e\, (2p)>-\beta^2\ .$

(iii) ${A}_{\pm}(\lambda, p)$ satisfy the functional
relation (so-called quantum Wronskian condition)
\eqn\wrons{e^{2\pi i p}\
A_{+}(q^{1\over 2}\lambda, p)\, A_{-}(q^{-{1\over 2}}\lambda, p)-
e^{-2\pi i p}\ A_{+}(q^{-{1\over 2}} \lambda, p)\,
A_{-}(q^{{1\over 2}}\lambda, p)=2 i\, \sin(2\pi p)\ ,}
where
\eqn\qdef{q=e^{i\pi \beta^2}.}

It is important to stress 
here that above conditions (i -- iii) define the
functions $A_{\pm}(\l,p)$ uniquely (see Appendix in Ref.\BLZnon).

Remarkably, according to\ \DT, in the special case $p = \beta^2/4$ 
the zeroes $\lambda_n^2$ 
of $A_{+}(\lambda, \beta^2/4)$ and $A_{-}(\lambda, \beta^2/4)$ 
coincide, up to an overall factor, with the
energy eigenvalues $E_n$ of the Schr\"odinger equation \anharm\ with
\eqn\abeta{
\alpha = {1\over\beta^2}-1\ .}
More precisely, let $E_n^-$ and $E_n^+$  ($n=1,2,3,\ldots$) be ordered 
eigenvalues corresponding to even and odd 
eigenfunctions in \anharm, respectively, and define the spectral
determinants \foot{Our notations for these functions slightly differ
from those in \DT\ -- the superscripts $\pm$ are interchanged and the
sign of the argument $E$ is reversed. In addition, there is a difference
in normalization -- ours corresponds to $D^{\pm}(0)=1$.},
\eqn\sdets{D^{\pm}(E)=\prod_{n=1}^{\infty}\Big(1-
{E\over E^{\pm}_n}\Big)\ .} 
The main statement of\ \DT\ is 
\eqn\AD{
A_\pm(\l,\beta^2/4)=D^{\pm}(\rho\l^2)\ , }
where
\eqn\Cval{
\rho=(2/\beta^2)^{2-2\beta^2}\,  \Gamma^2(1-\beta^2)\ .
}
The key part of the arguments in\ \DT\ leading to \AD\ is the functional   
relation which the spectral determinants \sdets\ obey; this relation 
was previously found in important series of works\ \refs{\Va,\Vb,\Vc}; 
it turns out to be identical to the functional relation \wrons.

We are going to show that a relation
similar to \AD\ holds for generic values
of $p$, if one replaces \anharm\ by the following more general spectral
problem. Consider the Schr\"odinger equation
\eqn\spd{ \partial_x^2 \Psi(x)+ \Big\{\, 
E-x^{2\alpha}-{l (l+1)\over x^2}\, \Big\}
\Psi(x)=0\ }
on the  half-line $0<x<\infty$. Here 
\eqn\lp{
l={{2p}\over\beta^2}-{1\over 2}
}
and again $\alpha$ is related to $\beta^2$ as in Eq.\abeta. Let us
assume that $\Re e\, l>-{3\over 2}$, and denote by $\psi(x,E, l)$ the
solution of \spd\ uniquely specified by the condition
\eqn\jsusdy{\psi(x,E,l):\ \ \ \psi(x,E, l)\to \sqrt{{2\pi\over 1+
\alpha}}\  \big(2+2\alpha\big)^{-{2l+1\over 2+2\alpha}}\ 
{x^{l+1}
\over \Gamma\big(1+{2l+1\over 2+2\alpha}\big)}+O(x^{l+3})\ \ \ \
{\rm as}\ \ \ \ \ \ x\to 0\ .  }
This solution can be analytically continued outside the domain 
$\Re e\, l>-{3\over 2}$. Obviously thus defined function 
$\psi(x,E,-l-1)$ solves the same equation \spd, and for generic
values of $l$ the solutions
\eqn\ksiduy{\psi^+(x,E)=\psi(x, E,l)\, ,\ \ \
\psi^-(x,E)=\psi^+(x,E,-l-1)\ ,}
are linearly independent, since 
\eqn\sksi{W\big[\psi^+,\psi^-\big]=2i\, \big(\, q^{l+{1\over 2}}-
q^{-l-{1\over 2}}\, \big)\ ,}
where $W\big[f, g\big]=f\, \partial_x g-g\, \partial_xf $ is the usual
Wronskian.
For certain isolated values of $E$ one of 
these solutions decays at $x\to +\infty$. Let 
$\big\{E_n^{+}\big\}_{n=1}^{\infty}$ and 
$\big\{E_n^{-}\big\}_{n=1}^{\infty}$ be ordered spectral sets defined 
by the conditions
\eqn\sidsuiu{\eqalign{&
\psi_n^+(x)\equiv\psi^{+}(x,E^{+}_n)\to 0\, , \cr
&\psi^{-}_n(x)\equiv\psi^{-}(x,E^{-}_n)\to 0\ ,}}
as\ $x\to+\infty$, \ 
and let
\eqn\ldets{D^{\pm}(E, l)=\prod_{n=1}^{\infty}\Big(1-
{E\over E^{\pm}_n}\Big)\ .}
Simple WKB analysis of the equation \spd\ shows that 
\eqn\eass{E^{\pm}_n\sim n^{{2\alpha\over 1+\alpha}}\ \ \ \  {\rm as}\
n\to\infty}
and therefore for $\alpha > 1$ these products converge, and 
\ldets\ defines entire functions of $E$. 
It is easy to see that in the special case $l=0$ the sets 
$\big\{E_n^{+}\big\}_{n=1}^{\infty}$ and 
$\big\{E_n^{-}\big\}_{n=1}^{\infty}$ become the components of the
spectrum of \anharm\ associated with odd and even sectors, respectively,
and so for $l=0$ the functions \ldets\ reduce to \sdets.
In what
follows we will show that for $\alpha>1$ and all values of $p$
\eqn\ADp{
A_{\pm}(\lambda,p) = D^{\pm}(\rho\l^2, 2p/\beta^2 -1/2) \ .}

We start with an observation that the 
following transformations of the variables $(x,
E, l)$, 
\eqn\sldoi{{\hat \Lambda}:\ x\to x\, , \ \ \
E\to E\, ,\ \ \  \
l\to-1-l\ ,}
\eqn\spod{{\hat \Omega}
:\ x\to q x \, ,\ \ \ E\to q^{-2}  E\,  ,\ \ \ \
l\to l}
with $q=e^{{i\pi\over 1+\alpha}}$, 
leave the equation \spd\ unchanged while acting nontrivially on its
solutions. As usual, the equation \spd\ admits
a unique solution which decays at large $x$; we denote this solution
as $\chi(x, E, l)$ and fix its normalization by the condition
\eqn\sjduuy{\chi(x,E, l) :\ \chi(x,E, l)\to x^{-{\alpha\over 2}}\
\exp\Big\{-{x^{1+\alpha}\over 1+\alpha}+O(x^{1-\alpha})\Big\}
\ \  \ {\rm as}\ x\to+\infty\ .}
The transformation \ ${\hat \Omega}$\ applied 
to\  $\chi(x, E, l)$\  yields
another solution, and the pair of functions
\eqn\sldi{\chi^+(x,E)=\chi(x, E, l)\, ,\ \ \  \ \
\chi^-(x,E)=i\, q^{-{1\over2}}\, \chi(qx,q^{-2} E, l)}
form a basis in the space of solutions of \spd. It is not difficult
to check that
\eqn\sksidse{
W\big[\chi^+,\chi^-\big]=2 \ ,}
i.e. the solutions \sldi\ are indeed linearly independent. The 
solutions \ksiduy\ can always be expanded in this basis, in particular
\eqn\skiuu{\psi^+=C(E,l)
\, \chi^++D(E,l)\,
\chi^-\ ,}
with some nonsingular coefficients $C(E,l)$ and $D(E,l)$. The
transformations \sldoi\ and \spod\ act on the solutions \ksiduy\ and 
\sldi\ as follows,
\eqn\ksudu{{\hat \Lambda}\psi^{\pm}=\psi^{\mp}\, ;\ \ \
{\hat \Lambda}\chi^{\pm}=\chi^{\pm}\ , }
\eqn\jsuy{{\hat \Omega}\psi^{\pm}=q^{1/2 \pm l\pm 1/2}\
\psi^{\pm}\, ;\ \ \
{\hat \Omega}\chi^{+}=
-i\, q^{1\over 2}\, \chi^{-}\, ,\ \ \ 
{\hat \Omega}\chi^{-}=-i\, q^{{1\over 2}}\, \chi^++u\, \chi^-}
with some  coefficient $u=u(E,l)$. It follows from \jsuy\ that
\eqn\sodi{C(E,l)=-i\ q^{-l-{1\over 2}}\ D(q^{-2} E,l)\ .  }
Also, applying \ksudu\  to\ \skiuu\  one obtains
\eqn\sdjuy{\psi^-=D(E,-l-1)\ 
\chi^--i\, q^{l+{1\over 2}}\, D(q^{-2} E,-l-1)\
\chi^+\ .}
Let us mention here a useful identity
\eqn\usefl{
D(E,l)={1\over 2}\ W\big[\chi^+ , \psi^+ \big]\ .
}

At this point we are ready to prove that the pair of functions $D(E,l)$
and
$D(E,-1-l)$ satisfy all the conditions (i -- iii) above and, since
these conditions characterize these functions uniquely,
\eqn\DA{
D(\rho\,\l^2,\, \pm 2p/\beta^2 -1/2) = A_{\pm}(\l, p)\ .}
Indeed, the analyticity conditions in (i) and (ii) can be derived from
\usefl, while asymptotics there are established by a
straightforward WKB analysis of the equation \spd. Finally, 
combining Eqs.$\skiuu,\, \sodi,\, \sdjuy,\, \sksi,\, \sksidse$\
one obtains the relation
\eqn\ikasiu{ q^{l+{1\over 2}}\, D(q^2E,l)\, D(E ,-l-1)-
q^{-l-{1\over 2}}\, D(E ,l)\, D(q^2 E,-l-1)= q^{l+{1\over 2}}-
q^{-l-{1\over 2}}\ ,  }
which is identical to \wrons.

To prove our statement \ADp\ it remains to show that the coefficient
$D(E,l)$ in \skiuu\ coincides with the function $D^{+}(E,l)$ defined in 
\ldets. Both are entire functions of $E$. As follows from \skiuu,
these functions share the same set of zeroes in the variable $E$ and
hence $F(E,l)=\log\big(D^{+}(E,l)/D(E,l)\big)$ is an entire function
of $E$. However,
$E\to\infty$ asymptotic form of $D^{+}(E,l)$ is 
controlled by asymptotic $n\to\infty$
density of the eigenvalues $E^{+}_n$ which can be computed
semiclassically. The result shows that $F(E,l)\to 0$ as $E\to\infty$ and
hence $F(E,l)=0$. 

Although strictly speaking our proof of \ADp\ 
is valid only if $\alpha > 1$, the
above arguments and the definition \ldets\ can be modified to accommodate
wider range of this parameter. We will not elaborate this point here.

In a few special cases the function ${A}_{+}(\lambda, p)$ was 
calculated explicitly \refs{\BLZZ, \Fen}. Let us consider these examples
to illustrate the identity \ADp. First, for harmonic oscillator case
$\alpha=1$ (which corresponds 
the $c=-2$ CFT, i.e. the ``free fermion'' theory) the spectrum of
\spd\ is very well known
\eqn\harm{
E^{+}_n=4n+2l-1\, ,\ \ \ \ n=1,\, 2\ldots\ ,}
which allows one to obtain
\eqn\skduiu{D^{+}(E,l)\big|_{\alpha=1}=
{\Gamma({3\over 4}+ {l\over 2})\ 
e^{{\cal C} E}\over
\Gamma({3\over 4}+ {l\over 2}-{E\over 4})}\ ,} 
where ${\cal C}$ is a constant whose value depends  on the choice of
Weierstrass factors required in this case to make the product
\ldets\ convergent.  The Eq.\skduiu\ is identical to the known
expression for $A_{+}(\l,p)\big|_{\beta^2 = 1/2}$ \BLZZ. 
Next, in the limit $\alpha\to+\infty$ and 
$l$ fixed (which corresponds to the classical limit $c\to-\infty$ in CFT)
the equation \spd\ reduces to the radial Schr\"odinger equation
for the spherically symmetric ``rigid well'' potential
\eqn\iwell{\eqalign{x^{2\alpha}\big|_{\alpha\to+\infty}=
\cases{0, & if $0<x<1$\cr
+\infty, &  if $x>1$\cr}\ ,  }} 
for which the  energy levels are related to the 
zeroes of the Bessel function, and \ldets\ yields
\eqn\besdetz{D^{+}(E, l)\big|_{\alpha\to+\infty}=
\Gamma(l+3/ 2)\  \big(\sqrt{E}/2\big)^{-l-{1\over 2}}\ 
J_{l+{1\over 2}}\big(\sqrt{E}\big)\ ,}
where $J_{\nu}(z)$ is the conventional Bessel function.
Again, this expression coincides with the limiting form of
$A_{+}(\l,p)\big|_{\beta^2\to 0}$ \BLZZ. Finally, for $\alpha=1/2$ and
$l=0$ the spectral 
determinants \ldets\ are expressed in terms of the Airy
function \refs{\Vc, \DT}, in agreement with $A_{\pm}(\l,1/6)\big|_{\beta^2
=2/3}$  obtained in \Fen. 

We would like to mention also some possible applications
of the relation\ \ADp.
In \BLZZ\ the  exact asymptotic expansions for  ${A}_{\pm}(\lambda, p)$
at large  $\lambda$ were found. The coefficients in these expansions 
are expressed in terms of the spectral characteristics of CFT,
namely the vacuum eigenvalues of its
local and non-local integrals of motion. In view of \ADp\ these
expansions can be
used to derive the large $n$ asymptotics of the energy
levels $E^\pm_n$.
In particular, the leading terms read
\eqn\syusd{\eqalign{ &E_n^{+}=\bigg[ {\sqrt{\pi} \Gamma\big({3\over 2}+
    {1\over 2\alpha}\big)\over2\, \Gamma\big(1+
    {1\over 2\alpha}\big)} \bigg]^{{2\alpha\over 1+\alpha}}   
    (4n+2l-1)^{{2\alpha\over 1+\alpha}}\times\cr & 
\ \ \ \ \ \     \bigg(\,1-{2\alpha\, \cot\big({\pi \over 2\alpha}\big)
\over 3\pi (1+\alpha)^2}\ { 12l^2+12 l-2\alpha+1\over
     (4n+2l-1)^2}+O\big(n^{-4}, n^{-1-2\alpha}
    \big)\, \bigg)\  \ \ \big(\, \alpha>{1\over 2}\, \big)\, ;\cr
&E_n^{+}=\bigg[ {\sqrt{\pi} \Gamma\big({3\over 2}+
    {1\over 2\alpha}\big)\over2\, \Gamma\big(1+
    {1\over 2\alpha}\big)} \bigg]^{{2\alpha\over 1+\alpha}}
    (4n+2l-1)^{{2\alpha\over 1+\alpha}}\
    \bigg(\, 1+
{2\alpha\, \Gamma\big(-{1\over 2}-\alpha\big)\over
(1+\alpha) \sqrt{\pi} \, \Gamma(-\alpha) }\times \cr &
 \bigg[ {2\, \Gamma\big(1+
    {1\over 2\alpha}\big)\over \sqrt{\pi} \Gamma\big({3\over 2}+
    {1\over 2\alpha}\big)}\bigg]^{2\alpha}\ 
  {\Gamma\big(l+\alpha+{3\over 2}\big)\over
  \Gamma\big(l-\alpha+{1\over 2}\big)}\  (4n+2l-1)^{-2\alpha-1}+
     O\big(n^{-2}, n^{-1-4\alpha}\big) \bigg)\,
\ \big(\, 0<\alpha<{1\over 2}\, \big)\, ;\cr
&E_n^{+}=\big(3\pi (4n+2l-1)\big)^{2\over 3}\ \bigg(\,
{1\over 4} +{12 l (l+1)\,  \log \big(3\pi (4n+2l-1)\big)\over
27\pi^2\, (4n+2l-1)^2}-\cr
&\ \ \ \ \ \ \ {18 l^2 +6 l-5+
12 l (l+1)\,  \psi(l+2)\over 27\pi^2\, (4n+2l-1)^2}+O\big(n^{-3}\big)
    \, \bigg)\ \ \ \big(\, \alpha={1\over 2}\, \big)\, .}}
where\ $\psi(z)=\partial_z \log \Gamma(z)$ and $n\to\infty$.
Using expressions for the local and non-local integrals of motion
given in \refs{\BLZZZZ,\BLZnon}, it is not difficult
to extend these expansions substantially further,
but first few  terms  already yield remarkable accuracy even for lower
levels $n\geq 4$.

The vacuum eigenvalues of the ${\bf Q}$-operators, i.e. the above
functions $A_{\pm}(\l,p)$, appear also  in studying a quantum problem of a
Brownian particle in the potential $U(X)=
-2\pi\kappa\,\cos(X)-VX$ at finite
temperature $T$. In
the Caldeira-Leggett 
approach\ \leggett\ this problem is related to so-called
boundary sine-Gordon model (with zero bulk mass)\ \Callan, 
and it was shown in\ \BLZnon\
that some expectation values in this problem are expressed through
the above functions $A_{\pm}(\l,p)$ with pure
imaginary $p\propto iV$ and
$\l \propto i\kappa$. In particular, the nonlinear mobility $J(V) =
\langle\,  {\dot X}\, \rangle$ is
\eqn\nlmob{
J(V) =V+ i\pi T\, \kappa\,\partial_\kappa\, \log \bigg[\, 
{{A_{+}(\sigma\kappa,
-i\,{V{\beta^2}\over{4\pi T}})}\over{A_{-}(\sigma\kappa,
-i\,{V{\beta^2}\over{4\pi T}})}}\, \bigg]\ ,  
}
where $\beta^2 \propto \hbar$ is a quantum parameter and $\sigma =
i\, \big(\, \beta^2/2\pi T\, \big)^{1-\beta^2}\,
\sin(\pi\beta^2)/\beta^2$ (see\ \BLZnon\ for
details). 
The equation \nlmob\ has nice interpretation in terms of the Schr\"odinger
problem \spd. To see this let us make a variable transformation in 
\spd, $x = e^y,\
\Psi = e^{y\over 2}\,{\tilde \Psi}$, which brings \spd\ to the
form 
\eqn\spdd{
-\partial_y^2\, 
{\tilde \Psi} + \Big\{\, e^{{2y\over\beta^2}}-E\,e^{2y}\, 
\Big\}\, {\tilde \Psi} =
\nu^2\, {\tilde \Psi}\, , \qquad \nu =i\, {{2p}\over \beta^2}\ .}
The potential term in \spdd\ decays at $y\to -\infty$ and therefore
for pure imaginary $p = -i\nu\beta^2/2$ \spdd\ defines 
a scattering problem, 
the reflection scattering amplitude
$S(\nu, E)$ being defined as usual as the coefficient in the $y\to
-\infty$ asymptotic
\eqn\sctass{
{\tilde \Psi}(y)
\to e^{i\nu y} + S(\nu, E)\,e^{-i\nu y} \qquad {\rm as} \qquad
y\to -\infty\ 
}
of the solution ${\tilde \Psi}(y)$ which decays at $y\to +\infty$. This
coefficient is readily extracted from $\skiuu,\,
\sodi$\ and \sdjuy,
\eqn\smat{
S(\nu, E) = -{\Gamma(1+i\nu\beta^2)
\over\Gamma(1-i\nu\beta^2)}\ {D(E,-{1\over
2}+i\nu)\over D(E,-{1\over
2}-i\nu)}\ \Big({\beta^2\over2}\Big)^{-2i\nu\beta^2}\  .}
Using the relations \DA\ we have for \nlmob\ 
\eqn\curr{
J(V) =V - 2i\pi  T\,
E\partial_E\,  \log S\Big({V\over2\pi T}\,,\, E\Big)\, ,\ \ \ \ 
E=-{\kappa^2\, \pi^2\over 
\Gamma^2(1+\beta^2)}\ \big(\pi T\big)^{2\beta^2-2}\ ,}
i.e. the nonlinear mobility \nlmob\ is expressed in terms of the
scattering phase in the Schr\"odinger problem \spdd.
Note that this expression allows one to prove the duality relation
$\beta^2\to{1/\beta^2}$
for $J(V)$, first proposed in Ref.\Fisher, by simple
change of the variables
in\ \spdd.

In conclusion, let us remark that the main ingredient used in derivation of
the functional equation \ikasiu\ is the symmetries $\sldoi,\, \spod$\ of
the Schr\"odinger operator \spd. Clearly, it is possible to modify this
operator while preserving these symmetries. This suggests that the
relation between the eigenvalues of ${\bf Q}$-operators of CFT 
and spectral
characteristics of Schr\"odinger operators may be generalizable to 
excited-states eigenvalues of ${\bf Q}_{\pm}$ and, more importantly, to
${\bf Q}$-operators corresponding to massive integrable field theories
(sine-Gordon model).

\bigskip

\centerline{ {\bf Acknowledgments}}
 
\bigskip
 
Research of S.L. and A.Z.
is supported in part by DOE grant \#DE-FG05-90ER40559.

\listrefs

\end